\RequirePackage{fix-cm}
\documentclass[smallextended]{svjour3}       
\smartqed  
\usepackage{graphicx}
\usepackage{graphicx}
\usepackage{amsmath}
\usepackage{amsfonts} 
\usepackage{amssymb}
\usepackage{longtable}
\newcommand{\be}{\begin{equation}}
\newcommand{\ee}{\end{equation}}

\newcommand{\ba}[1]{\left(\begin{array}{#1}}
\newcommand{\ea}{\end{array}\right)}
\journalname{International Journal of Theoretical Physics}
\begin{document}
\title{Sum Uncertainty Relations: Uncertainty Regions for Qubits and
Qutrits} \thanks{A.R.Usha Devi and Sudha acknowledge financial support from the Department of Science and Technology, India through Project No. DST/ICPS/QuST/Theme-2/2019/Project#107}


\titlerunning{Uncertainty Regions for Qubits and Qutrits}        

\author{Seeta Vasudevrao \and I. Reena \and Sudha \and A. R.
Usha Devi  \and A. K. Rajagopal}         

\institute{Seeta Vasudevrao \at
              Department of Physics, Bangalore University, Bangalore. 
       \email{yergolkarseeta@gmail.com}           
           \and I. Reena
            \at Department of Physics, Bangalore University, Bangalore. 
\email{reena.irevinod24@gmail.com} 
\and
Sudha
\at Department of Physics, Kuvempu University, Shankaraghatta. \\
              Inspire Institute Inc., Alexandria, Virginia, 22303, USA.
							\email{arss@rediffmail.com} 
\and
A. R. Usha Devi
\at Department of Physics, Bangalore University, Bangalore \\
Inspire Institute Inc., Alexandria, Virginia, 22303, USA.
\email{arutth@rediffmail.com} 
\and 
A. K. Rajagopal 
\at  Inspire Institute Inc., Alexandria, Virginia, 22303, USA.	
\email{attipat.rajagopal@gmail.com}}

\maketitle
\begin{abstract}
We investigate the notion of {\em uncertainty region} using the  variance based sum uncertainty relation for qubits and qutrits. We compare {\em uncertainty region} of  the qubit (a 2-level system) with that of the qutrit (3-level system) by considering sum uncertainty relation for two non-commuting Pauli-like observables, acting  on the two dimensional qubit Hilbert space. We identify that physically valid uncertainty region of a qubit is smaller than that of a qutrit. This implies that an enhanced precision can be achieved in the measurement of incompatible Pauli-like observables acting on the 2-dimensional subspace of a qutrit Hilbert space.  We discuss the implication of the reduced uncertainties in the steady states of $\Lambda$, V, $\Xi$ types of 3-level atomic systems. Furthermore, we construct a  two-qubit permutation symmetric state, corresponding to a 3-level system and show that the reduction in the sum uncertainty value - or equivalently, increased uncertainty region of a qutrit system -- is  a consequence of quantum entanglement in the two-qubit system. Our results suggest that uncertainty region can be used as a dimensional witness.         

\keywords{Sum Uncertainty Relation \and Uncertainty Region \and 
Entanglement \and 3-Level System \and Steady-state Population}
\end{abstract}
\section{Introduction}
\label{intro} 
Quantum theory prevents assignment of precise values for two or more incompatible observables simultaneously. Heisenberg's heuristic argument \cite{Heisenberg} highlighted this uncertainty associated with the non-commuting
position ($Q$) and momentum ($P$) observables, in terms of the constraints placed
on the product of their standard deviations. If position of a particle is measured,
prediction of its momentum gets inaccurate and vice versa. A mathematically formal version
of the position-momentum uncertainty relation 
$(\Delta Q) (\Delta P) \geq  \frac{\hbar}{2}$ was subsequently formulated by Kennard~\cite{ken}. Furthermore, Robertson~\cite{rob} (motivated
by Weyl's arguments~\cite{Weyl}) extended the uncertainty relations to any arbitrary
pairs of non-commuting observables $A_1$, $A_2$. 

Different forms
of uncertainty relations have been formulated over the years~\cite{AK,ozawa,Hall,BLW,Hoffman,sw1,arun,kraus,MU,sw,Berta,Coles,Werner,Li,Abbot}, capturing
the trade-off between two or more non-commuting observables. It has been
shown that uncertainty relations play a crucial role in quantum information
processing tasks like quantum key distribution~\cite{Berta,Coles,Cerf1,Cerf2,Koshi}. Non-trivial state-dependent uncertainty relations which are experimentally verifiable are found to be of importance in device-independent cryptography~\cite{sw1}.

A broader perspective on uncertainty relations, based on the concept of
{\emph {uncertainty regions}}, is recently being explored~\cite{Werner,Li,Abbot,Busch} and provides a geometric visualization of the uncertainty relation. For any uncertainty relation, the corresponding uncertainty region
is the {\emph {legitimate domain}} of standard deviation (or any other measure
of uncertainties) of a pair (or triple) of observables, in the entire range of
their possible values~\cite{Busch}. Points $(\Delta A_1,\, \Delta A_2)$ inside the uncertainty region specify the uncertainty in the simultaneous measurement of a pair of observables $A_1$, $A_2$.  
Different types of uncertainty relations can be chosen for analysing the  uncertainty regions~\cite{Werner,Li,Abbot,Busch}. In this work we have chosen variance based sum uncertainty relation, a state-independent uncertainty relation,  proposed by Hofmann and Takeuchi~\cite{Hoffman}, to analyze the uncertainty regions/minimum of the variance based sum uncertainty relation for two non-commuting Pauli-like observables. 

The structure of the paper is as follows: In Section~2, we outline the geometry of uncertainty region corresponding to variance based sum-uncertainty relation for a pair of incompatible Pauli-like  observables, when they are measured in quantum states of  qubits (2-level systems) and qutrits (3-level systems).  We show that
uncertainty region of qutrits is larger, containing points with enhanced measurement precision for incompatible Pauli-like observables, in comparison with that of qubits. 
In Section~3, we express the minimum of the sum of variances of two noncommuting Pauli-like observables $A^{(ij)}_1=\vec{\sigma}^{(ij)}\cdot\hat{a}$,\ $A^{(ij)}_2=\vec{\sigma}^{(ij)}\cdot\hat{b},\ \ \hat{a}\cdot\hat{b}=0$  of a 3-level system, given that both the observables are restricted to the 2-dimensional (qubit) subspace (labelled by the pair of indices $(ij),\,  i< j=1,2,3$)  of a 3-level (qutrit) system,  in terms of the populations $\rho_{ii}$, $\rho_{jj}$  in the  $i^{\rm th}$ and $j^{\rm th}$ levels.  We discuss implication of the reduced uncertainties in the steady states of $\Lambda$, V and $\Xi$ types of 3-level atomic systems. 
Section~4 details the construction of permutation symmetric two-qubit system  corresponding to a qutrit state and explicit evaluation of equivalent sum uncertainty relation for Pauli-like observables. In Section~5, we establish that separable two-qubit states can never achieve utmost precision in the measurement of incompatible Pauli-like observables. We also show that maximum precision in measurement of the non-commuting
Pauli observables is possible using entangled symmetric two-qubit states. 
Section 6 provides concluding remarks. 

\section{Uncertainty regions for qubits and qutrits:} 
The well-known generalized uncertainty relation~\cite{rob} for observables $A_1$, $A_2$ is given by 
\be
\label{rs}
(\Delta A_1) (\Delta A_2) \geq \frac{1}{2} \left \vert  \langle [A_1,\, A_2 ] \rangle \right \vert
\ee
where $[A_1,\,A_2]=A_1A_2-A_2A_1$ is the commutator and $\Delta A_1$, $\Delta A_2$ defined by
\begin{eqnarray}
\Delta A_1&=&\sqrt{\Delta^2\,A_1},\ \ \ \ \ \Delta^2 A_1=\langle A^2_1 \rangle-\langle A_1 \rangle^2 \nonumber \\
\Delta B&=&\sqrt{\Delta^2\,A_2},\ \ \ \ \ \Delta^2 A_2=\langle A^2_2 \rangle-\langle A_2 \rangle^2.
\end{eqnarray}
are the standard deviations of $A_1$, $A_2$ in any quantum state $\rho$. Here,
$\langle \cdots \rangle= \mbox{Tr}\,(\rho\cdots)$ is the expectation value of any observable in the state $\rho$.

Hofmann and Takeuchi~\cite{Hoffman} reformulated the uncertainty relation   (\ref{rs}) in
the form of sum of variances. Given a set of non-commuting operators $A_i$, $i=1,\,2,\,\ldots,\ n$, they have shown that 
\be
\label{aika}
\sum_{i=1}^n\,\Delta^2\,A_i \geq \, k_A, \ \ k_A \ \mbox{being a non-negative real number.} 
\ee 
It is a {\emph{state-independent}} uncertainty relation~\cite{Hoffman} with non-trivial bound for incompatible 
observables.  

We now set up the sum-uncertainty relation in  (\ref{aika})  for a pair of observables $A_1$, $A_2$ acting on the most general state of a qubit: 
\be
\label{qb1}
\rho_{\rm qubit}=\frac{1}{2} \left[I_2+r_1\sigma_1+r_2\sigma_2+r_3\sigma_3 \right],\ \ \ r_1^2+r_2^2+r_3^2 \leq 1
\ee
where $\sigma_i$, $i = 1,\, 2,\,3$ are Pauli spin operators, $I_2$ is the two-dimensional identity 
operator and $\vec{r} = (r_1,\,r_2,\, r_3)$, ($\vert\vec{r}\vert\leq 1$) is a real three dimensional vector, the mean spin vector of the state $\rho_{\rm qubit}$. On choosing $A_1=\sigma_1$ and $A_2=\sigma_2$, we get 
$\langle A_1^2 \rangle=\langle A_2^2 \rangle=1$, $\langle A_1 \rangle=r_1$, $\langle A_2 \rangle=r_2$. With $\Delta^2\,A_1=1-r_1^2$, 
$\Delta^2\,A_2=1-r_2^2$, the sum-uncertainty relation becomes
\begin{eqnarray} 
& & \Delta^2\,A_1+\Delta^2\,A_2=2-(r_1^2+r_2^2)\geq 1, \\ 
& & 0\leq \Delta A_1\leq 1,\ \ \ \ \ \ 0\leq \Delta A_2\leq 1. \nonumber
\end{eqnarray}
In general, we consider the {\emph{orthogonal}} Pauli-observables
\be
\label{pauli}
A_1=\vec{\sigma}\cdot \hat{a},\ \ \  A_2=\vec{\sigma}\cdot \hat{b}, \ \ \vec{\sigma}=\left(\sigma_1,\,\sigma_2,\,\sigma_3 \right),\ \ \ \hat{a}\cdot \hat{b}=0
\ee 
and we readily have $\langle A_1\rangle=\hat{a}\cdot \vec{r}$, 
$\langle A_2\rangle=\hat{b}\cdot \vec{r}$, $\langle A_1^2\rangle=\hat{a}\cdot \hat{a}=1$, 
$\langle A_2^2\rangle=\hat{b}\cdot \hat{b}=1$ leading to 
\begin{eqnarray}
\label{int}
& & \Delta^2\,A_1=1-\left(\hat{a}\cdot \vec{r}\right)^2,\ \ \ \ \Delta^2\,A_2=1-\left(\hat{b}\cdot \vec{r}\right)^2 \nonumber \\ 
& & \Delta^2\,A_1+\Delta^2\,A_2=2-\left(\hat{a}\cdot \vec{r}\right)^2-\left(\hat{b}\cdot \vec{r}\right)^2.
\end{eqnarray}
As $\vert\vec{r}\vert\leq 1$, $\vert \hat{a} \vert=\vert \hat{b} \vert=1$, we have $\hat{a}\cdot \vec{r}\leq 1, \ \  
\hat{b}\cdot \vec{r}\leq 1$ and from Eq. (\ref{int}) we obtain the following sum-uncertainty relation for orthogonal 
Pauli-observables on a qubit: 
\be
\label{qbur1}
\Delta^2\,A_1+\Delta^2\,A_2\geq 1\ \ \ \mbox{with}\ \ \ 
0\leq \Delta A_1\leq 1,\ \ \ 0\leq \Delta A_2\leq 1.
\ee 
The sum uncertainty relation (\ref{qbur1}) implies that the points $(\Delta A_1,\,\Delta A_2)$ lying outside the circular quadrant form the uncertainty region for Pauli-observables $A_1$, $A_2$ measured 
on a qubit, as can be seen in Fig.~1. 
\begin{figure}
\includegraphics{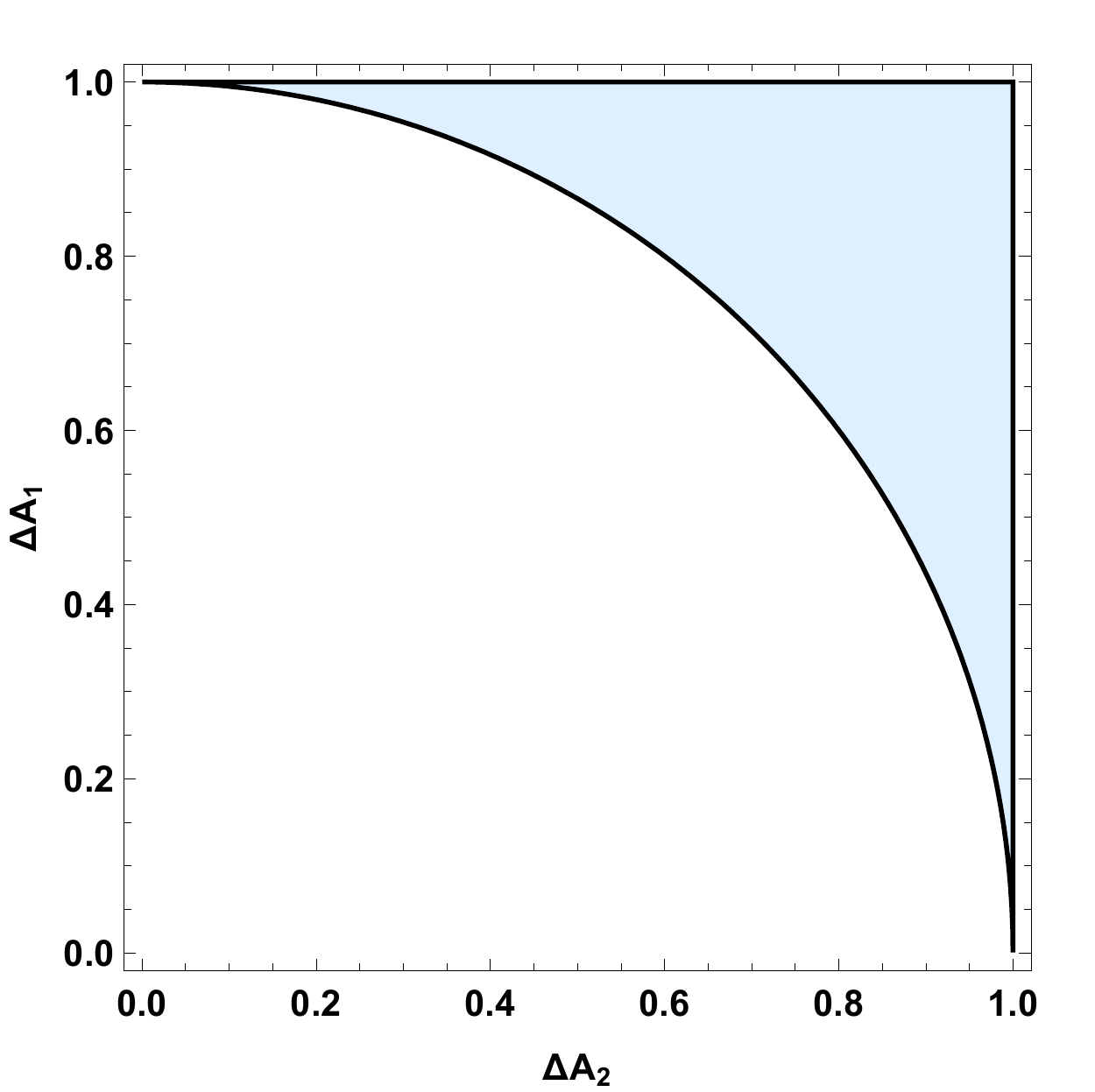}
\caption{The uncertainty region (light-blue shaded) of a qubit for orthogonal Pauli observables $A_1$, $A_2$ in  (\ref{pauli}):}
\label{one}       
\end{figure} 

Notice that the points $(\Delta A_1,\,\Delta A_2)$ close to the origin correspond to measurements that result in better accuracy. But as the uncertainty region does not contain points below the circular arc (See Fig.~1), precise joint measurements of Pauli-observables on a qubit are impossible and Fig.~1 provides clear visualization of this fact.


We now consider a 3-level system,
\be
\label{qt1}
\rho_{\rm qutrit}=\omega \vert \psi\rangle\langle \psi \vert \oplus (1-\omega),\ \ 0\leq \omega \leq 1,  
\ee
obtained by appending an ancillary level to a single-qubit pure state $\vert \psi \rangle$.  Here (\ref{qt1}) corresponds to the state of a
qutrit whose explicit form is given by  
\be
\label{qt2a}
\rho_{\rm qutrit}=\ba{ccc} \frac{\omega(1+r_3)}{2} & \frac{\omega(r_1-ir_2)}{2} & 0  \\ 
\frac{\omega(r_1+ir_2)}{2} & \frac{\omega(1-r_3)}{2}  & 0  \\
 0 & 0 &  1-\omega  \ea, \ \ \ r_1^2+r_2^2+r_3^2=1
\ee
Here, $r_1$, $r_2$, $r_3$ are the components of the {\emph {unit mean spin vector}} $\hat{r}$, corresponding to the {\emph{pure state}} $\vert \psi\rangle$ (See  (\ref{qt1})) and $\omega$ is a real parameter. 
 
The uncertainty region of the qutrit in   (\ref{qt1}) has been examined in Ref.~\cite{Busch}, for orthogonal Pauli observables
\be
\label{qtob}
A_1=\vec{\sigma}\cdot \hat{a}\oplus 0,\ \ \ \ A_2=\vec{\sigma}\cdot \hat{b}\oplus 0, \ \ \hat{a}\cdot \hat{b}=0.
\ee 
It can be seen that~\cite{Busch} 
\be
\label{qt2}
\langle A_1 \rangle=\omega(\hat{a}\cdot \hat{r}),\ \ \ \langle A_2 \rangle=\omega(\hat{b}\cdot \hat{r}), \ \ 
\langle A_1^2\rangle=\langle A_2^2\rangle=\omega,
\ee
leading to 
\be
\label{qt3}
\Delta^2 A_1=\omega-\omega^2 (\hat{a}\cdot \hat{r})^2,\ \ \ \Delta^2 A_2=\omega-\omega^2 (\hat{b}\cdot \hat{r})^2.
\ee 
On fixing  $\Delta^2 A_1$ and minimizing $\Delta^2 A_2$, one obtains~\cite{Busch}
\be
\label{qt4}
\left(\Delta A_2\right)_{\rm min}=\Delta A_1\sqrt{(1-\Delta^2 A_1)}.
\ee 
Similarly,  fixing $\Delta^2 A_2$ and minimizing $\Delta^2 A_1$ results in~\cite{Busch} 
\be
\label{qt5}
\left(\Delta A_1\right)_{\rm min}=\Delta A_2\sqrt{(1-\Delta^2 A_2)}.
\ee 
From  (\ref{qt4}), (\ref{qt5}), we see that when $\Delta A_1=0$, $\Delta A_2$ can also become zero and vice
versa. This means, origin $(0,\, 0)$ of the co-ordinate system, a point corresponding to utmost precision in simultaneous measurement, is  physically realizable for measurement of orthogonal Pauli observables on the qutrit state $\rho_{\rm qutrit}$~\cite{Busch}. The uncertainty region
of the qutrit is larger in comparison with that of a qubit, containing points near the origin and origin itself, as can be readily
seen in Fig.~2. 
\begin{figure}
\includegraphics{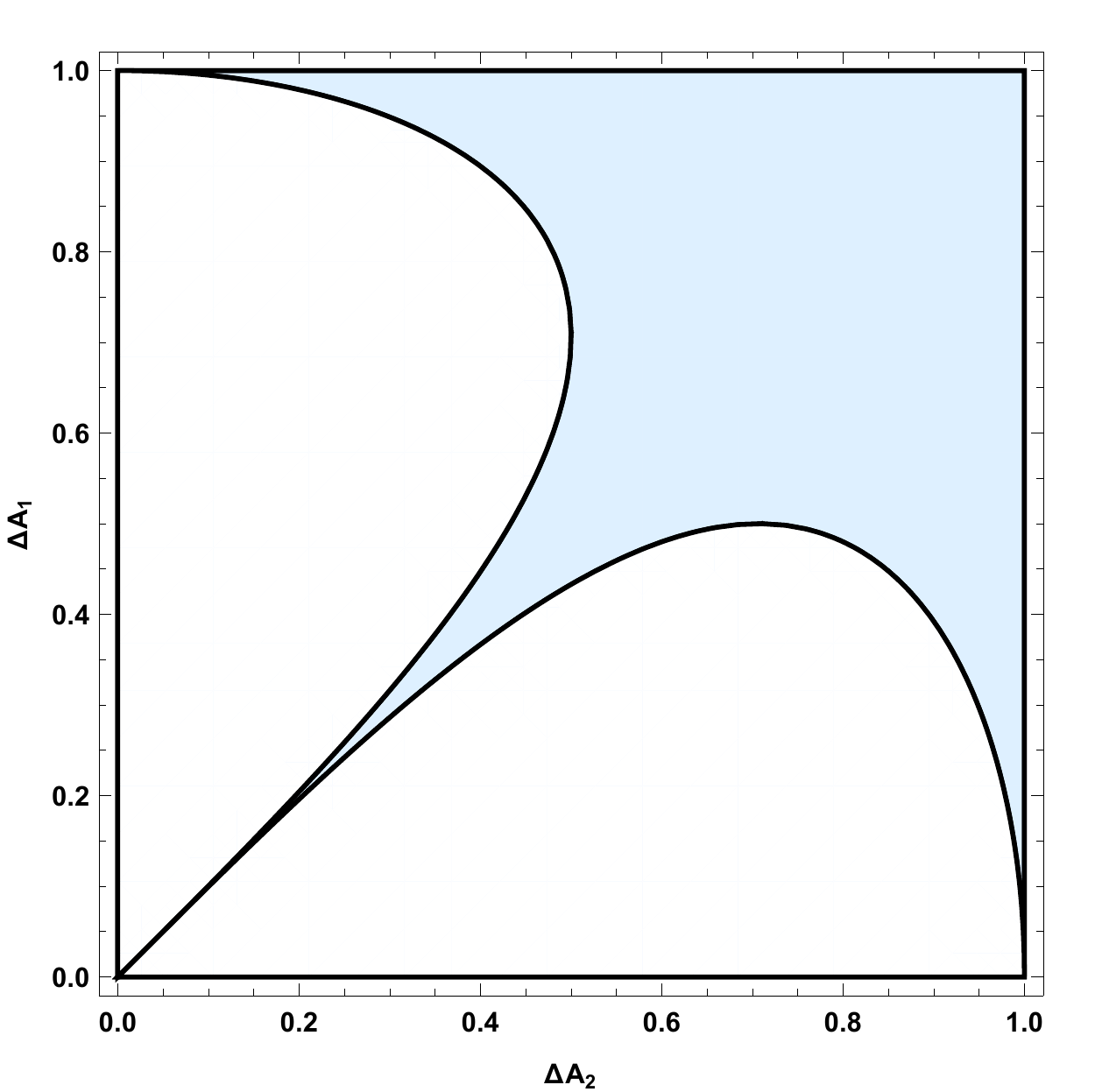}
\caption{The uncertainty region of a qutrit (light blue):  The origin, with $\Delta A_1 = \Delta A_2 =0$ and points closer to it lie in the admissible region.}
\label{two}       
\end{figure}

It is evident from the above observations that simultaneous measurements with utmost precision are impossible when non-commuting Pauli measurements are performed on a 2-dimensional Hilbert space (qubit) whereas a 3-dimensional Hilbert space (qutrit) admits enhanced precision. In the following, we show that the uncertainty sum for two Pauli-like observables in an arbitrary 3-level system reduces below that for a 2-level system.

\section{Sum uncertainty relation for 3-level atomic systems}

In this section we explore the sum uncertainty relation for two Pauli-like observables (i.e., atomic operators acting on any 2-level subspace of a 3-level atomic system)  
\begin{eqnarray}
\label{aij}
A^{(ij)}_1&=&\vec{\sigma}^{(ij)}\cdot\hat{a}, \ \  \vec{\sigma}^{(ij)}=\left(\sigma^{(ij)}_1,\,\sigma^{(ij)}_2,\,\sigma^{(ij)}_3 \right) \nonumber \\ 
A^{(ij)}_2&=&\vec{\sigma}^{(ij)}\cdot\hat{b},\ \  \ \   i<j=1,\,2,\,3,
\end{eqnarray}  
where $\hat{a}\cdot\hat{b}=0$, $\hat{a}\cdot\hat{a}=1=\hat{b}\cdot\hat{b}$ and  \ \
\begin{eqnarray*} 
	\sigma^{(12)}_1&=& \left(\begin{array}{ccc} 0 & 1 & 0 \\ 1 & 0 & 0\\  0 & 0 & 0 \end{array}\right), \ \ \ \ \sigma^{(13)}_1 = \left(\begin{array}{ccc}0 & 0 & 1 \\ 0 & 0 & 0\\  1 & 0 & 0\end{array}\right), \ \ \  \sigma^{(23)}_1 = \left(\begin{array}{ccc}0 & 0 & 0 \\ 0 & 0 & 1\\  0 & 1 & 0\end{array}\right) \\  
	\sigma^{(12)}_2&=&\left(\begin{array}{ccc}0 & -i & 0 \\ i & 0 & 0\\  0 & 0 & 0\end{array}\right), \ \ \sigma^{(13)}_2 = \left(\begin{array}{ccc}0 & 0 & -i \\ 0 & 0 & 0\\  i & 0 & 0\end{array}\right), \ \ \sigma^{(23)}_2 = \left(\begin{array}{ccc}0 & 0 & 0 \\ 0 & 0 & -i\\  0 & i & 0\end{array}\right) \\
	\sigma^{(12)}_3&=& \left(\begin{array}{ccc}1 & 0 & 0 \\ 0 & -1 & 0\\  0 & 0 & 0\end{array}\right), \ \ \sigma^{(13)}_3 = \left(\begin{array}{ccc}1 & 0 & 0 \\ 0 & 0 & 0\\  0 & 0 & -1\end{array}\right), \ \ \sigma^{(23)}_3 = \left(\begin{array}{ccc}0 & 0 & 0 \\ 0 & 1 & 0\\  0 & 0 & -1\end{array}\right).    
\end{eqnarray*}
In an arbitrary 3-level atomic system, characterized by the density matrix,  
\begin{equation}
\label{varrho}
\varrho_{{\rm qutrit}}=\left(\begin{array}{ccc} \varrho_{11} & \varrho_{12} & \varrho_{13} \\  
\varrho^*_{12} & \varrho_{22} & \varrho_{23} \\
\varrho^*_{13} & \varrho^*_{23} & \varrho_{33} \\
\end{array}\right),
\end{equation}
we obtain
\begin{eqnarray}
\label{aijsq1}
\left\langle \left(A_1^{(ij)}\right)^2\right\rangle &=& {\rm Tr}\,\left[\varrho_{\rm qutrit}\,\left(\vec{\sigma}^{(ij)}\cdot\hat{a}\right)^2\right]=\varrho_{ii}+\varrho_{jj}, \\ 
\label{aijsq2}
\left\langle \left(A_2^{(ij)}\right)^2\right\rangle &=& {\rm Tr}\,\left[\varrho_{\rm qutrit}\,\left(\vec{\sigma}^{(ij)}\cdot\hat{b}\right)^2\right]=\varrho_{ii}+\varrho_{jj},                           
\end{eqnarray}
and 
\begin{eqnarray}
\label{aijmean1}
\left\langle A_1^{(ij)}\right\rangle &=& {\rm Tr}\,\left[\varrho_{\rm qutrit}\,\left(\vec{\sigma}^{(ij)}\cdot\hat{a}\right)\right]=\vec{n}^{(ij)}\cdot\hat{a}, \\ 
\label{aijmean2}
\left\langle A_1^{(ij)}\right\rangle &=& {\rm Tr}\,\left[\varrho_{\rm qutrit}\,\left(\vec{\sigma}^{(ij)}\cdot\hat{b}\right)\right]=\vec{n}^{(ij)}\cdot\hat{b}, 
\end{eqnarray}
where 
\begin{eqnarray}
\label{nij}
\vec{n}^{(ij)}={\rm Tr}\,[\varrho_{\rm qutrit}\, \vec{\sigma}^{(ij)}]= \left( 2\,{\rm Re}\,\varrho_{ij},\, 2\, {\rm Im}\,\varrho_{ij},\, \varrho_{ii}-\varrho_{jj}\right).
\end{eqnarray}
Choosing $\hat{a}=\hat{n}^{(ij)}=\vec{n}^{(ij)}/\vert\vec{n}^{(ij)}\vert$, $\hat{b}=\hat{n}^{(ij)}_{\perp}$ and  simplifying the sum of variances $\Delta^2 A_1^{(ij)}+\Delta^2 A_2^{(ij)}$   in the 3-level system (\ref{varrho})  (with the help of  (\ref{aijsq1}), (\ref{aijsq2}), (\ref{aijmean1}), (\ref{aijmean2}), (\ref{nij}))), we obtain    
\begin{eqnarray}
\label{uncpgen}
\left[\Delta^2 A_1^{(ij)}+\Delta^2 A_2^{(ij)}\right]&=& 2\,(\varrho_{ii}+\varrho_{jj}) - \vert\vec{n}^{(ij)}\vert^2 \nonumber \\
&=& 2\,(\varrho_{ii}+\varrho_{jj})-\left[4\, \vert \varrho_{ij} \vert^2+(\varrho_{ii}-\varrho_{jj})^2\right].
\end{eqnarray}
Positive semidefiniteness of the  $(ij)^{\rm th}$ $2\times 2$ block of $\varrho_{\rm qutrit}$ imposes the condition 
\be
\vert\vec{n}^{(ij)}\vert\leq \varrho_{ii}+\varrho_{jj}
\ee
leading to the following minimum value for the uncertainty sum: 
\begin{eqnarray}
\label{uncpmin}
\left[\Delta^2 A_1^{(ij)}+\Delta^2 A_2^{(ij)}\right]_{\rm min}&=& 2\,\left(\varrho_{ii}+\varrho_{jj}\right) - \left(\varrho_{ii}+\varrho_{jj}\right)^2. 
\end{eqnarray}
It may be noted that if  we   restrict ourselves to the  2-level system i.e., $i,j=1,2$, we obtain     
$\left[\Delta^2 A_1^{(12)}+\Delta^2 A_2^{(12)}\right]_{\rm min}=1$, as $(\varrho_{11}+\varrho_{22})={\rm Tr}[\varrho]=1$. In other words, the uncertainty sum is always greater than 1 in a 2-level system indicating that joint measurement of the non-commuting Pauli observables $\vec{\sigma}\cdot\hat{a},\ \vec{\sigma}\cdot\hat{b}$ is limited by the sum uncertainty relation (\ref{qbur1}). On the other hand, when an additional level is included, the  populations  $\varrho_{ii}$, $\varrho_{jj}$ of the  $i^{\rm th}$ and $j^{\rm th}$ levels ($i>j=1,2,3$) play a crucial role in enhancing the measurement precision of the atomic observables $\vec{\sigma}^{(ij)}\cdot\hat{a}$, \ $\vec{\sigma}^{(ij)}\cdot\hat{b}$. 

In Fig.~3 we have plotted the minimum value of the uncertainty sum $\left[\Delta^2 A_1^{(ij)}+\Delta^2 A_2^{(ij)}\right]_{\rm min}$  as a function of the populations $\varrho_{ii}$, $\varrho_{jj}$ (see (\ref{uncpmin})). It is clearly seen  that the minimum value of the uncertainty sum (\ref{uncpmin}) can take values smaller than 1. This establishes the advantage of the additional level for improving  precision in the measurement of  atomic observables.          
\begin{figure}
	\label{3level}
	\includegraphics{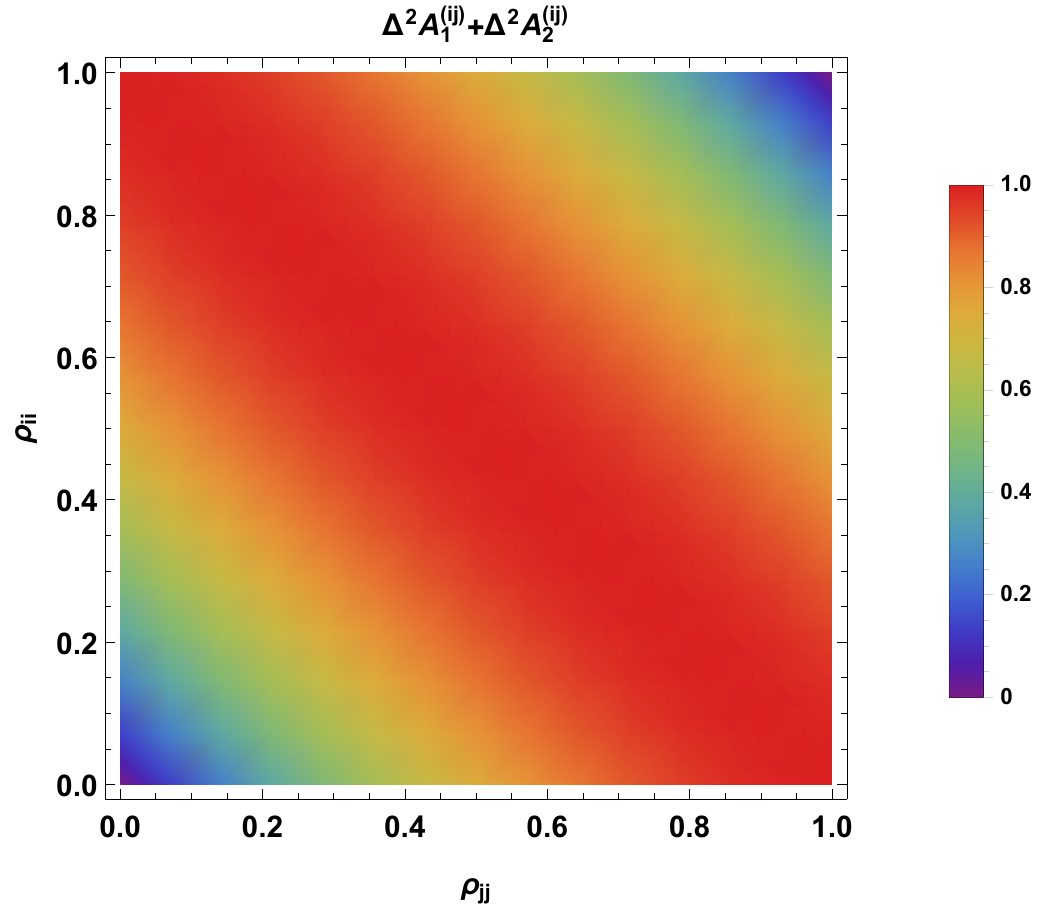}
	\caption{Minimum value of the uncertainty sum $\left[\Delta^2 A_1^{(ij)}+\Delta^2 A_2^{(ij)}\right]_{\rm min}$   
		as a function of the populations $\rho_{ii}$, $\rho_{jj}$ (see (\ref{uncpmin})). Enhanced measurement precision of incompatible Pauli-like atomic observables  
		is ensured whenever $\left[\Delta^2 A_1^{(ij)}+\Delta^2 A_2^{(ij)}\right]_{\rm min}<1$.}
	\label{3}       
\end{figure}

Considerable research interest has been evinced in exploring the response of $\Lambda$, $\Xi$, V types of 3-level atomic systems to lasing radiation~\cite{3levelReview}. In $\Lambda$-type system (see Fig.~4(a)) transition between the two lower levels $\vert 3\rangle$, $\vert 2\rangle$ is  forbidden  and the upper level $\vert 1\rangle$ is commonly shared in atomic transitions with levels $\vert 3\rangle$ and $\vert 2\rangle$; in V-type system (Fig.~4(b)), transitions from the lower level  $\vert 3\rangle$  with the two upper levels $\vert 1 \rangle$ and $\vert 2\rangle$ are allowed, but the transition $\vert 1\rangle\leftrightarrow\vert 2\rangle$ between the upper levels is forbidden. While atomic transitions  $\vert 1\rangle\leftrightarrow\vert 2\rangle$   and $\vert 2\rangle\leftrightarrow\vert 3\rangle$ are allowed in   $\Xi$ type system (Fig.~4(c)),  the transition  $\vert 1\rangle\leftrightarrow\vert 3\rangle$  is forbidden. 
It is of interest to consider any two levels of the atomic 3-level system, between which  transitions are allowed, as a qubit,  and explore if there is any enhanced precision in the measurement of two non-commuting qubit operators. To this end, we consider  the coherent population trapping state in a $\Lambda$  type atomic system~\cite{radmore82,pra95,pro_opt96,pra97}: 
\begin{eqnarray}
\varrho^{\Lambda}_{33}=\frac{1}{2}=\varrho^{\Lambda}_{22},\ \  \varrho^{\Lambda}_{11}=0. 
\end{eqnarray}

It is clearly seen that the uncertainty sum  (\ref{uncpmin})  with  $i=1,\, j=2$ and $i=1,\, j=3$ is given by  
\begin{eqnarray*}
\left[\Delta^2 A_1^{(12)}+\Delta^2 A_2^{(12)}\right]^\Lambda_{\rm min}=2\, (\varrho^\Lambda_{11}+\varrho^\Lambda_{22})-(\varrho^\Lambda_{11}+\varrho^\Lambda_{22})^2=0.75  \\
	\left[\Delta^2 A_1^{(13)}+\Delta^2 A_2^{(13)}\right]^\Lambda_{\rm min}=2\, (\varrho^\Lambda_{11}+\varrho^\Lambda_{33})-(\varrho^\Lambda_{11}+\varrho^\Lambda_{33})^2=0.75
	\end{eqnarray*}
revealing improved precision in the measurements of the qubit operator pairs $\left\{A_1^{(12)},\, A_2^{(12)}\right\}$ and  $\left\{A_1^{(13)},\, A_2^{(13)}\right\}$. 

In the case of V-type 3-level atom, with the transition $\vert 3\rangle\leftrightarrow \vert 2\rangle$ driven by a strong-coupling laser field and $\vert 3\rangle\leftrightarrow \vert 1\rangle$ transition  driven by an incoherent pump field the steady state populations are given by~\cite{pra96}  
\begin{eqnarray}
\label{vss}
\varrho^{V}_{11}\approx 0.2,\ \varrho^{V}_{22}\approx\varrho^{V}_{33}\approx 0.4. 
\end{eqnarray}
The uncertainty sum (\ref{uncpmin})  of Pauli-like atomic observables associated with   $i=1,\, j=3$ and $i=2,\, j=3$ are given by    
\begin{eqnarray*}
	\left[\Delta^2 A_1^{(13)}+\Delta^2 A_2^{(13)}\right]^V_{\rm min}=2\, (\varrho^V_{11}+\varrho^V_{33})-(\varrho^V_{11}+\varrho^V_{33})^2=0.84  \\
	\left[\Delta^2 A_1^{(23)}+\Delta^2 A_2^{(23)}\right]^V_{\rm min}=2\, (\varrho^V_{22}+\varrho^V_{33})-(\varrho^V_{22}+\varrho^V_{33})^2=0.96.
\end{eqnarray*} 
Thus  the steady state of V-type atomic qutrit (see (\ref{vss})) offers advantage over 2-level atomic system in reducing the uncertainty sum of non-commuting Pauli-like atomic observables. 
\begin{center}                
	\begin{figure}
		\includegraphics[scale=0.37]{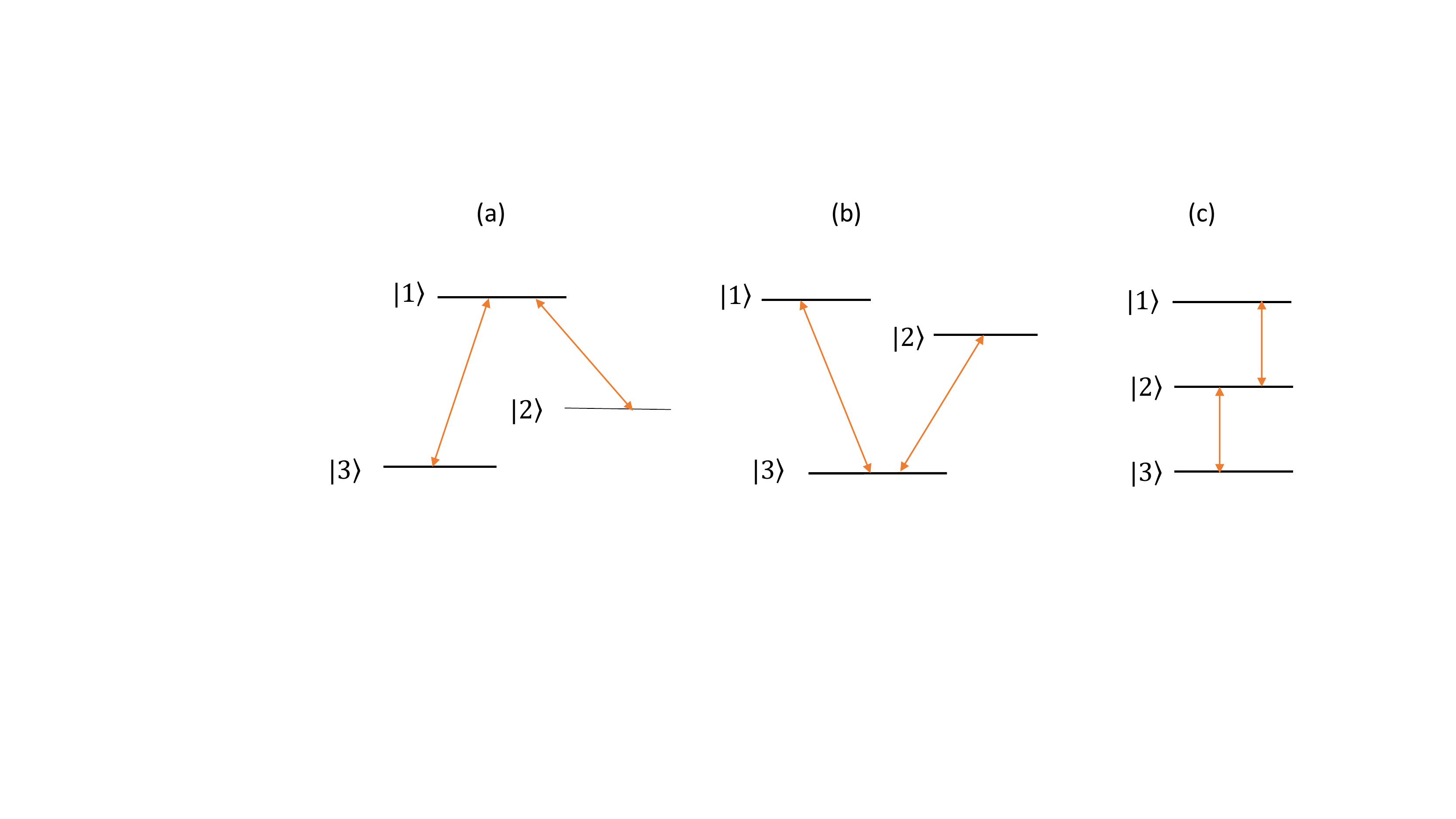}
		\caption{Schematic diagrams of  (a) $\Lambda$-type (b) V-type and (c) $\Xi$-type 3-level  atomic systems. Arrows between the energy levels indicate allowed transitions.}
		\label{3}       
	\end{figure} 
\end{center}

Populations in the steady state of a 3-level $\Xi$ atomic system~\cite{pra97,pra98} satisfy the condition
\begin{eqnarray}
\varrho^{\Xi}_{11}=\varrho^{\Xi}_{22}\leq \frac{1}{3},\ \ \  \frac{1}{3}\leq \varrho^{\Xi}_{33}\leq \frac{1}{2}.
\end{eqnarray} 
We thus obtain  the limiting value of the uncertainty sum as,  
\begin{eqnarray*}
	\left[\Delta^2 A_1^{(12)}+\Delta^2 A_2^{(12)}\right]^\Xi_{\rm min}&=&2\,\left(1-\varrho^\Xi_{33}\right)-\,\left(1-\varrho^\Xi_{33}\right)^2; \ \ \   \frac{1}{3}\leq \varrho^{\Xi}_{33}\leq \frac{1}{2},  \nonumber \\ 
	&\implies& \hskip 0.2in \frac{3}{4} \leq \left[\Delta^2 A_1^{(12)}+\Delta^2 A_2^{(12)}\right]^\Xi_{\rm min}\leq \frac{8}{9}, \\ 
	 \left[\Delta^2 A_1^{(23)}+\Delta^2 A_2^{(23)}\right]^\Xi_{\rm min}&=&2\,\left(1-\varrho^\Xi_{11}\right)-\,\left(1-\varrho^\Xi_{11}\right)^2;\ \ \  0\leq \varrho^{\Xi}_{11}\leq \frac{1}{3} ,   \nonumber \\ 
	 &\implies& \hskip 0.2in  \frac{8}{9} \leq \left[\Delta^2 A_1^{(12)}+\Delta^2 A_2^{(12)}\right]^\Xi_{\rm min}\leq 1, \\ 
\end{eqnarray*}
for the atomic Pauli-like observables associated with  $i=1,\, j=2$ and $i=2, j=3$ atomic levels of  the $\Xi$-type atomic system. 
More detailed investigations on improving measurement precision of non-commuting Pauli-like atomic observables of a driven 3-level system, beyond what can be achieved in a 
2-level system, will be reported separately.

\section{Transformation of qutrit state into a two-qubit symmetric state} 

We now analyze the enhanced accuracy of orthogonal Pauli measurements on qutrit states from a perspective based on the separability/non-separability of two-qubit symmetric states corresponding to qutrit states.  In the following, we detail the construction of 
two-qubit {\emph{symmetric states}} from qutrit states, in particular the state in  (\ref{qt1}). We also carry out an analysis of the role, if any, of two-qubit entanglement in the measurement precision possible in the qutrit state (\ref{qt1}).

A two-qubit symmetric state belongs to the 3-dimensional maximum multiplicity space 
of the collective angular momentum $j = j_1 + j_2$, $j_1 = j_2 = \frac{1}{2}$. 
With the dimensions of the qutrit space and that of the two-qubit symmetric
state (expressed in angular momentum basis) being equal, they have a one-one
correspondence. Here, we outline this correspondence and accomplish the construction of two-qubit symmetric states corresponding to  qutrit states. 

The qutrit state in  (\ref{qt1}), expressed as a  $3\times 3$ matrix in  (\ref{qt2a}), can be written equivalently
as
\be
\label{qtmtx}
\rho_{\rm qutrit}\equiv\ba{cccc} \frac{\omega(1+r_3)}{2} & \frac{\omega(r_1-ir_2)}{2} & 0 & 0 \\ 
\frac{\omega(r_1+ir_2)}{2} & \frac{\omega(1-r_3)}{2}  & 0 & 0 \\
 0 & 0 &  1-\omega  & 0 \\ 
0 & 0 & 0 & 0
     \ea, 
\ee 
The density matrix $\rho_{AB}$ of a two-qubit symmetric state is given by~\cite{uma}
\be
\label{2qbt}
\rho_{AB}=\frac{1}{4}\left[I_2\otimes I_2+\sum_{i=1}^3\,s_i\left(\sigma_i\otimes I_2+
I_2\otimes \sigma_i\right)+\sum_{i,j=1}^3\, t_{ij}\, (\sigma_i\otimes \sigma_j)   \right],  
\ee
where $t_{ij}=t_{ji}$. 
The elements $\rho_{ij}$, $i,\,j=1,\,2,\,3,\,4$ of the $4\times 4$ matrix $\rho_{AB}$ are explicitly given by 
\begin{eqnarray}
\label{elements} 
\rho_{11}&=&\frac{1}{4}\left(1+2s_3+t_{33}\right), \ \ \  \rho_{22}=\frac{1}{4}\left(1-t_{33}\right)=\rho_{33} \nonumber \\  
& & \nonumber \\
\rho_{12}&=&\rho_{13}=\frac{1}{4}\left(s_1-is_2+t_{13}-it_{23}\right)=\rho^\ast_{21}=\rho^\ast_{31} \nonumber \\ 
& & \nonumber \\
\rho_{14}&=&\frac{1}{4}\left(t_{11}-t_{22}-2it_{12}\right)=\rho^\ast_{14} \nonumber \\ 
& & \nonumber \\
\rho_{24}&=&\rho_{34}=\frac{1}{4}\left(s_1-is_2-t_{13}+it_{23}\right)=\rho^\ast_{42}=
\rho^\ast_{43} \nonumber \\ 
& & \nonumber \\
\rho_{23}&=&\frac{1}{4}\left(t_{11}+t_{22}\right)=\rho_{32}, \ \ \  \rho_{44}=\frac{1}{4}\left(1-2s_3+t_{33}\right) 
\end{eqnarray}
The standard basis for a two-qubit state is given by the direct product basis (uncoupled basis) 
consisting of orthonormal vectors
\be
\vert 1/2;1/2\rangle, \ \ \vert 1/2;-1/2\rangle,\ \ \vert -1/2;1/2\rangle, \ \ \vert -1/2;-1/2\rangle 
\ee
where $\vert m_1;m_2\rangle=\vert m_1\rangle\otimes m_2\rangle$, $m_1,\,m_2=1/2,\,-1/2$,

One can readily express the direct product basis $\{\vert m_1;m_2\rangle\}$ in terms of the collective angular momentum basis (coupled basis) 
 $\{\vert jm\rangle\}$, ($j=1,\,0$, $-j\leq m \leq j$ for each $j$) and vice versa, through 
\begin{eqnarray}
\label{cuc}
\vert 11\rangle&=&\vert 1/2;1/2\rangle,\ \ \ \vert 10\rangle=\frac{1}{\sqrt{2}}\left(\vert 1/2;-1/2\rangle+\vert -1/2;1/2\rangle\right) \\
\vert 1-1\rangle&=&\vert -1/2;-1/2\rangle,\ \ \ \vert 00\rangle=\frac{1}{\sqrt{2}}\left(\vert 1/2;-1/2\rangle-\vert -1/2;1/2\rangle\right). \nonumber
\end{eqnarray}
From  (\ref{cuc}), it follows that the unitary matrix 
\be
\label{unitary}
U=\ba{cccc} 1 & 0 & 0 & 0 \\ 0 & 0 & 0 & 1  \\ 0 & \frac{1}{\sqrt{2}}  & \frac{1}{\sqrt{2}} & 0 \\   
0 & \frac{1}{\sqrt{2}} & \frac{-1}{\sqrt{2}} & 0 \ea, \ \ U^\dagger\,U=U\,U^\dagger=I_4 
\ee
where $^\dagger$ denotes hermitian conjugate, corresponds to the transformation from coupled basis to uncoupled basis.  The similarity transformation $U^\dagger \rho_{\rm qutrit} U$ effects the transformation of the state $\rho_{\rm qutrit}$ (See  (\ref{qt1}),  (\ref{qtmtx})) to its equivalent 
two-qubit symmetric state $\rho_{AB}$. Explicitly, we have 
\be
\label{qtmatrixff}
\rho_{AB}=U^\dagger \rho_{\rm qutrit} U =\frac{1}{2}\ba{cccc} (1+r_3)\omega & 0 & 0 & (r_1-ir_2)\omega \\ 
0 & 1-\omega & 1-\omega  & 0  \\ 0 & 1-\omega & 1-\omega  & 0 \\ (r_1+ir_2)\omega  & 0 & 0 & (1-r_3)\omega  \ea.
\ee
On comparing the elements (See  (\ref{elements})) of $\rho_{AB}$  with the corresponding elements in Eq. (\ref{qtmatrixff}), we obtain the following relation between the parameters $\omega$, $r_1$, $r_2$, $r_3$, 
($r_1^2+r_2^2+r_3^2=1$) of the qutrit state in  (\ref{qt1}) and the parameters $s_i$, $t_{ij}$, $i,\,j=1,\,2,\,3$  of the symmetric two-qubit state $\rho_{AB}$ in  (\ref{2qbt}).  
That is, the non-zero parameters $s_i$, $t_{ij}$, $i,\,j=1,\,2,\,3$ of $\rho_{AB}$ (See  (\ref{2qbt})) are seen to be 
\begin{eqnarray}
\label{stow}
s_3&=&\omega r_3,\ \ \ t_{12}=t_{21}=\omega r_2    \\ 
t_{11}&=&(1-\omega)+\omega r_1,\ \ \ \ t_{22}=(1-\omega)-\omega r_1,\ \ t_{33}=2\omega-1. \nonumber  
\end{eqnarray}
Thus, a two-qubit symmetric state $\rho_{AB}$ in  (\ref{2qbt})) with its elements given in  (\ref{stow})
corresponds to the qutrit state $\rho_{\rm qutrit}$ in  (\ref{qt1}). 

\section{Sum uncertainty relation for two-qubit state $\rho_{AB}$} 

Here, we set up the sum uncertainty relation for
the two-qubit state $\rho_{AB}$ (See  (\ref{qtmatrixff})) in order to establish the equivalence of its uncertainty region with that of the qutrit state $\rho_{\rm qutrit}$ (See  (\ref{qt1})).
In order to do this, we need to recognize the two-qubit observables ${\mathcal A}_1$, ${\mathcal A}_2$,  which are equivalent to  $A_1$, $A_2$ (See  (\ref{qtob})).
For simplicity, and without loss of generality, we choose the orthonormal vectors in  (\ref{qtob}) to be $\hat{a} = (1,\, 0,\, 0)$, 
$\hat{b} = (0,\, 1,\, 0)$ so that  $A_1=\sigma_1 \oplus 0$, $A_2=\sigma_2 \oplus 0$. It is not difficult to see that we can express $A_1$, $A_2$ as 
\be
\label{a1a2qt}
A_1=\sigma_1\oplus {\bf 0}_2, \ \  A_2=\sigma_2\oplus {\bf 0}_2,\  \   {\bf 0}_2=\ba{cc} 0 & 0 \\ 0 & 0 \ea, 
\ee
to facilitate their action on the qutrit state $\rho_{\rm qutrit}$ expressed as a $4\times 4$ matrix in  (\ref{qtmtx}). Corresponding to the basis transformation $\rho_{\rm qutrit}\longrightarrow \rho_{AB}$ in  (\ref{qtmatrixff}), the observables $A_1$, $A_2$ 
(See  (\ref{a1a2qt})) undergo the similarity transformation
\begin{eqnarray}
\label{a1toa}
{\mathcal A}_1&=&U^\dag\,A_1 U=\frac{1}{2} \left[ \sigma_1\otimes \sigma_1- \sigma_2\otimes \sigma_2 \right], \nonumber \\ 
{\mathcal A}_2&=&U^\dag\,A_2 U=\frac{1}{2} \left[ \sigma_1\otimes \sigma_2+ \sigma_2\otimes \sigma_1 \right].
\end{eqnarray}
Here $U$ is the basis transformation matrix (See (\ref{unitary})) that takes $\rho_{\rm qutrit}$  to $\rho_{AB}$ (See  (\ref{qtmatrixff})).   On explicit evaluation, we get 
\begin{eqnarray}
\label{sum2qbt}
&&\langle {\mathcal A}_1 \rangle=\mbox{Tr}\,({\mathcal A}_1\rho_{AB})=\omega\,r_1,\ \ \ \langle {\mathcal A}_1^2 \rangle=\mbox{Tr}\,({\mathcal A}_1^2\rho_{AB})=\omega \nonumber \\ 
&&\langle {\mathcal A}_2 \rangle=\mbox{Tr}\,({\mathcal A}_2\rho_{AB})=\omega\,r_2,\ \ \ \langle {\mathcal A}_2^2 \rangle=\mbox{Tr}\,({\mathcal A}_2^2\rho_{AB})=\omega \nonumber \\ 
&&\Delta^2 {\mathcal A}_1=\omega-\omega^2\,r_1^2,\ \ \ \ \ \ \ \ \ \ \ \Delta^2 {\mathcal A}_2=\omega-\omega^2\,r_2^2.
\end{eqnarray} 
The expressions for $\Delta^2 {\mathcal A}_1$, $\Delta^2 {\mathcal A}_2$ in  
 (\ref{sum2qbt}) are the same as that  obtained in  (\ref{qt3}) for the qutrit state $\rho_{\rm qutrit}$ (See (\ref{qt1})). 
The uncertainty region of the two-qubit state $\rho_{AB}$ is thus
the same as that of the qutrit state $\rho_{\rm qutrit}$, with the origin $(\Delta {\mathcal A}_1,\, \Delta {\mathcal A}_2)=(0,\,0)$  
being a physically realizable point (See Fig.~2). 

As $r_1^2+r_2^2+r_3^2=1$, the sum uncertainty relation of the two-qubit state $\rho_{AB}$ can be simplified to (See  (\ref{sum2qbt}))
\be
\Delta^2 {\mathcal A}_1+\Delta^2 {\mathcal A}_2=2\omega-\omega^2\,\kappa^2,\ \ \ \kappa=\sqrt{r_1^2+r_2^2}=\sqrt{1-r_3^2}.
\ee
Fig.~4 shows the variation of the uncertainty sum $\Delta^2 {\mathcal A}_1+\Delta^2 {\mathcal A}_2$ as a function of the parameters $0\leq \omega,\kappa\leq 1$. 
\begin{figure}
\includegraphics{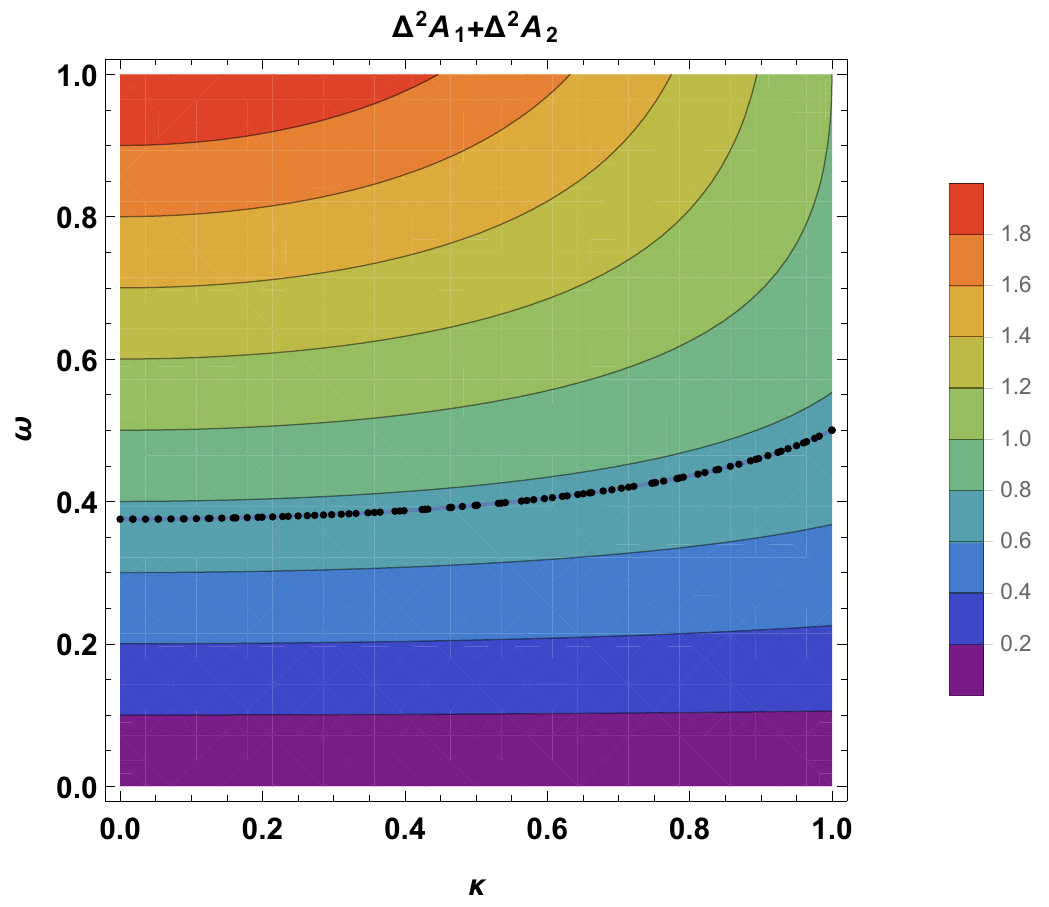}
\caption{The contour plot showing the variation of the uncertainty sum $\Delta^2 {\mathcal A}_1+\Delta^2 {\mathcal A}_2$ 
with respect to the parameters $\omega$ and $\kappa$. The dotted line corresponds to 
$\Delta^2 {\mathcal A}_1+\Delta^2 {\mathcal A}_2=\frac{3}{4}$.}
\label{3}       
\end{figure} 

\subsection{Sum uncertainty relation for symmetric two-qubit separable states} 
Our intention in obtaining the two-qubit counterpart $\rho_{AB}$ of the qutrit state $\rho_{\rm qutrit}$ lies in utilizing its {\emph {composite nature}} and examine whether separability/non-separability of $\rho_{AB}$ has any role in the better precision observed in joint measurement of Pauli observables on the {\emph{single}} party state $\rho_{\rm qutrit}$.  In order to carry out this task, we consider the most general bipartite, symmetric separable 
state  
\be
\label{rhosep}
\rho_{\rm sep}=\sum_{i}\,p_i \left( \rho_i\otimes \rho_i\right), \ \ i=1,\,2,\,3\ldots
\ee
with $0\leq p_i\leq 1$, $\sum_i\,p_i=1$ being the probabilities. 
It has been shown in Ref.~\cite{pfun} that the single qubit density operators $\rho_i$, $i=1,\,2,\,3\ldots$ constituting any symmetric separable state $\rho_{\rm sep}$ are necessarily {\emph {pure}}. Thus,  
\be
\label{qusep}
\rho_i=\frac{1}{2}\, \left(I_2+\vec{\sigma}\cdot \hat{s}_i \right), \ \  \hat{s}_i=\left(s_{1i},\,s_{2i},\,s_{3i} \right), \ \ 
s_{1i}^2+s_{2i}^2+s_{3i}^2=1. 
\ee
We evaluate the expectation values of ${\mathcal A}_\alpha,$ and  ${\mathcal A}^2_\alpha$,  $\alpha=1,2$ (See (\ref{a1toa})) in a product state $\rho_i\otimes \rho_i$ (where $\rho_i$ is given by   (\ref{qusep})):  
\begin{eqnarray*}
\langle {\mathcal A}_1 \rangle_i&=&\mbox{Tr}\,\left[\left( \rho_i\otimes \rho_i\right){\mathcal A}_1\right]=\frac{1}{2}\, \left(s_{1i}^2-s_{2i}^2\right), \nonumber \\
\langle {\mathcal A}_1^2 \rangle_i&=&\mbox{Tr}\,\left[\left( \rho_i\otimes \rho_i\right){\mathcal A}_1^2\right]=\frac{1}{2}\, \left(1+s_{3i}^2\right), \nonumber \\
\langle {\mathcal A}_2 \rangle_i&=&\mbox{Tr}\,\left[\left( \rho_i\otimes \rho_i\right){\mathcal A}_2\right]=s_{1i}s_{2i}, \nonumber \\
\langle {\mathcal A}_2^2 \rangle_i&=&\mbox{Tr}\,\left[\left( \rho_i\otimes \rho_i\right){\mathcal A}_2^2\right]=\frac{1}{2}\, \left(1+s_{3i}^2\right), \nonumber
\end{eqnarray*} 
leading to 
\begin{eqnarray}
\left(\Delta^2\,{\mathcal A}_1\right)_i&=&\frac{1}{2}\left[1+s_{3i}^2-\frac{1}{2}(s_{1i}^2-s_{2i}^2)^2 \right] \nonumber \\ 
\left(\Delta^2\,{\mathcal A}_2\right)_i&=&\frac{1}{2}\left[1+s_{3i}^2-2s_{1i}^2s_{2i}^2 \right]  
\end{eqnarray}
Using $s_{1i}^2+s_{2i}^2+s_{3i}^2=1$ and on simplification,  we get 
\be
\label{funcsum}
\left(\Delta^2\,{\mathcal A}_1\right)_i+\left(\Delta^2\,{\mathcal A}_2\right)_i=\frac{3}{4}+\frac{3}{2}s_{3i}^2-
\frac{1}{4}s_{3i}^4
\ee
From the structure of $\rho_{\rm{sep}}$ (See  (\ref{rhosep})), we readily have $\Delta^2 {\mathcal A}_1=\sum_{i}\,p_i\,\left(\Delta^2\,{\mathcal A}_1\right)_i$, $\Delta^2 {\mathcal A}_2=\sum_{i}\,p_i\,\left(\Delta^2\,{\mathcal A}_2\right)_i$ and hence we get (See  (\ref{funcsum})) 
\be
\Delta^2\,{\mathcal A}_1+\Delta^2\,{\mathcal A}_2=\frac{3}{4}+\frac{3}{2} \sum_i\,p_i\,s_{3i}^2-
\frac{1}{4}\sum_i\,p_i\,s_{3i}^4
\ee
As $0\leq s_{3i}\leq 1$, it readily follows that 
\be
\left(\Delta^2\,{\mathcal A}_1+\Delta^2\,{\mathcal A}_2\right)_{\rm min}=\frac{3}{4}
\ee
which happens when $s_{3i}=0$ for all $i=1,\,2,\,3,\cdots$. In other words, 
\be
\label{sepsum}
\Delta^2\,{\mathcal A}_1+\Delta^2\,{\mathcal A}_2 \geq \frac{3}{4}
\ee
is the sum-uncertainty relation for symmetric separable two-qubit states, with its lowest bound being $3/4$. Thus the uncertainty sum $\rho_{\rm sep}$ (See  (\ref{rhosep})),  set up for the two-qubit observables ${\mathcal A}_1$, ${\mathcal A}_2$ in  (\ref{a1toa}) cannot even go close to zero. This implies that symmetric separable states (See (\ref{rhosep})) can never achieve maximum accuracy in  joint measurements by ${\mathcal A}_1$, ${\mathcal A}_2$ in  (\ref{a1toa}). 

We now wish to check whether entanglement in the two-qubit state $\rho_{AB}$ contributes to enhanced precision in the   measurements of the observables ${\mathcal A}_1$, ${\mathcal A}_2$. To this end, we  
evaluate the concurrence~\cite{Wootters,hill}, a measure of two-qubit entanglement, of the state $\rho_{AB}$. Concurrence of any arbitrary two-qubit state $\rho$ is defined as~\cite{Wootters,hill}
\be
C=\mbox{max}\,\left(0,\,\sqrt{\lambda_1}-\sqrt{\lambda_2}-\sqrt{\lambda_3}-\sqrt{\lambda_4} \right)
\ee
where $\lambda_k$, $k=1,\,2,\,3,\,4$ are the eigenvalues of the matrix $\rho(\sigma_y\otimes\sigma_y)\rho^\ast(\sigma_y\otimes\sigma_y)$, arranged in the descending order (i.e., $\lambda_1\geq \lambda_2\geq \lambda_3\geq \lambda_4$). 

The structure of $\rho_{AB}$ in   (\ref{qtmatrixff})  allows us to make use of the simplified expression for
concurrence given in Ref.~\cite{wang}, and leads to  
\be
C_{AB}=\bigg \{
\begin{array}{cc}
\omega(1+\kappa)-1 \ \ \ \mbox{for} \ \ \ \omega(1+\kappa)\geq 1  \\
 1-\omega(1+\kappa) \ \ \ \mbox{for} \ \ \ \ \omega(1+\kappa)\leq 1 
\end{array}
\ee
where $\kappa=\sqrt{1-r_3^2}$ and $0\leq r_3\leq 1$. In other words, we have  
\be 
C_{AB}=\left\vert \omega(1+\kappa)-1 \right\vert, \ \ \kappa=\sqrt{1-r_3^2},\ \ 0\leq r_3\leq 1.
\ee 
A contour plot of $C_{AB}$ as a function of the parameters $\omega$ and $\kappa$ is shown in Fig.~6.

\begin{figure}
\includegraphics{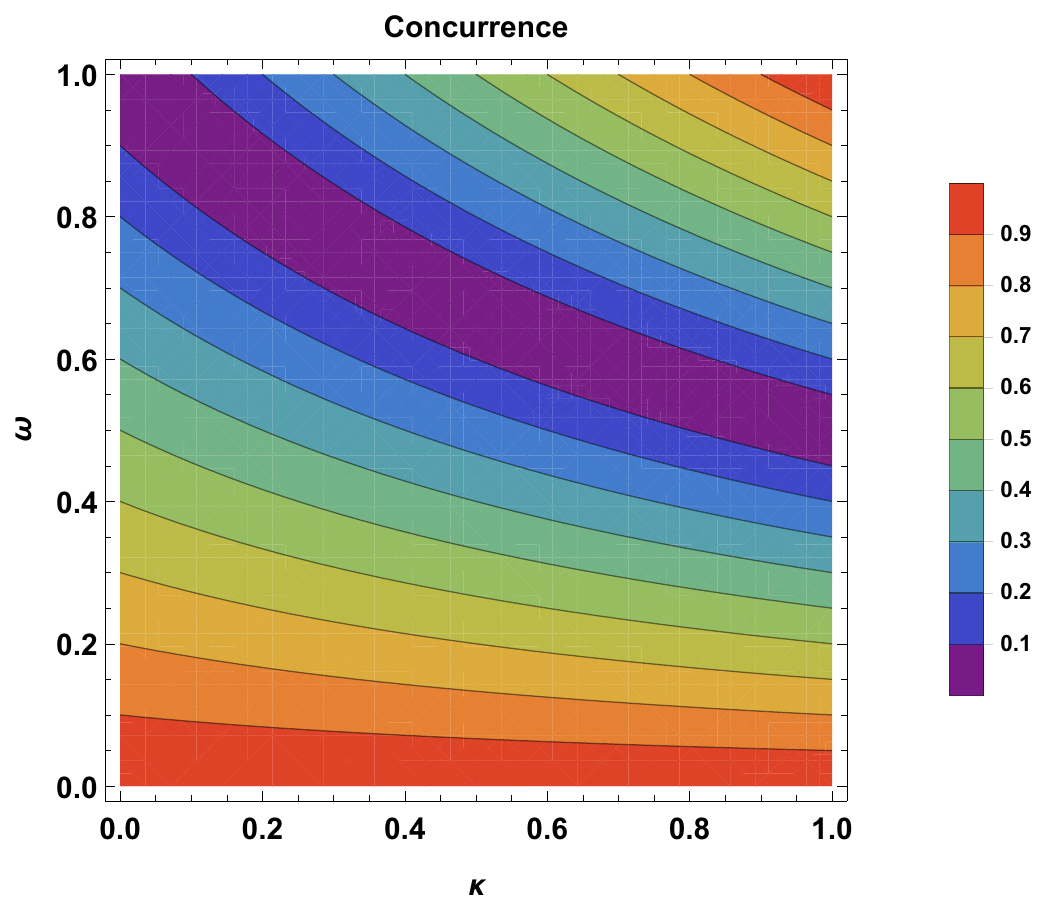}
\caption{The contour plot showing the variation of concurrence $C_{AB}$
with respect to the parameters $\omega$ and $\kappa$}
\label{6}       
\end{figure}

Based on  Figs.~5 and 6, we reach the following conclusions:  
\begin{enumerate}
\item  The uncertainty sum  $\Delta^2\,{\mathcal A}_1+\Delta^2\,{\mathcal A}_2$ can be reduced below the 
value $3/4$ only in an entangled state $\rho_{AB}$  (i.e., when $C_{AB}\neq 0$). In particular, 
$\Delta^2\,{\mathcal A}_1+\Delta^2\,{\mathcal A}_2\longrightarrow 0$, in maximally entangled two-qubit symmetric states (i.e., when  $C_{AB}\rightarrow 1$). 
\item While no separable state can reduce the uncertainty sum $\Delta^2\,{\mathcal A}_1+\Delta^2\,{\mathcal A}_2$  below the value $3/4$, there indeed exist entangled states with   $\Delta^2\,{\mathcal A}_1+\Delta^2\,{\mathcal A}_2\geq 3/4$. 
This implies that, while qutrit states $\rho_{\rm qutrit}$ that permit accurate simultaneous measurements of the observables  $A_1,\ A_2$ are necessarily associated with entangled two-qubit states, the converse is not always true. 
\end{enumerate}

It can thus be concluded that entanglement in a two-qubit state constructed from a qutrit (a qubit appended with an additional level) plays a significant role in the precise joint measurements by a pair of orthogonal Pauli observables. It would be of interest to examine whether two-qubit states constructed from a qudit (a qubit with $d-1$ ancillary levels) exhibit a similar feature. A study of uncertainty region of such two-qubit states, dimensional dependence of uncertainty sum and accuracy of simultaneous measurements by {\emph {any}} incompatible pair of observables in  $d$-dimensional spaces form topics of further interest.

\section{Conclusion} 
This work is a contribution to the ongoing study on uncertainty regions~\cite{Werner,Li,Abbot,Busch} providing a different perspective in accounting for better measurement precision seen in qutrits.  For any arbitrary 3-level atomic systems we have obtained an expression for minimum value of uncertainty sum for Pauli-like observables  in terms of atomic populations. This is useful to study if enhanced measurement precision (reduction in the uncertainty sum) can be realized in $\Lambda$, V and $\Xi$ types of  3-level atomic systems, which are characterised by  different schemes of allowed/forbidden atomic transitions between any two levels.
We have also examined whether entanglement in a two-qubit state (which is constructed from a qutrit state of specific structure) is responsible for the inclusion of more precisely determinable points in the uncertainty region. We have shown that while simultaneous measurements of orthogonal Pauli-like observables with utmost accuracy is not possible for separable two-qubit states, entangled states allow for such precise
measurements.  The technique we have proposed for the construction of two-qubit symmetric states corresponding to qutrit states is helpful in such a construction from multi-level states obtained by appending several ancillary levels to a single qubit. It would be of interest to examine whether increase in the dimension of Hilbert space leads to better measurement precision of incompatible observables. Our work suggests that uncertainty region could be employed as a dimensional witnesse.


 \end{document}